\documentclass[a4paper]{jpconf}
\usepackage{graphicx}
\usepackage{amssymb}
\usepackage{tikz,slashed}
\newcommand{\beq}{\begin{equation}}
\newcommand{\eeq}{\end{equation}}
\newcommand{\bea}{\begin{eqnarray}}
\newcommand{\eea}{\end{eqnarray}} 
\newcommand{\beqa}{\begin{eqnarray}}
\newcommand{\eeqa}{\end{eqnarray}} 

\newcommand{\bit}{\begin{itemize}}
\newcommand{\eit}{\end{itemize}}

\newcommand{\bfi}{\begin{figure}}
\newcommand{\efi}{\end{figure}}

\newcommand{\ben}{\begin{enumerate}}
\newcommand{\een}{\end{enumerate}}
\newcommand{\bbk}{\begin{block}}
\newcommand{\ebk}{\end{block}}
\newcommand{\bex}{\begin{example}}
\newcommand{\eex}{\end{example}}
\newcommand{\ave}[1]{{\langle{#1}\rangle}}
\newcommand{\ket}[1]{| {#1} \rangle}
\newcommand{\bra}[1]{\langle {#1} |}

\begin{document}
\title{Baryon-rich QCD matter}

\author{Jochen Wambach}

\address{Institut f{\"u}r Kernphysik\\
Theoriezentrum\\

Technische Universit{\"a}t Darmstadt}

\ead{wambach@physik.tu-darmstadt.de}

\begin{abstract}
Properties of high-density strong-interaction matter of relevance for astrophysical scenarios that involve neutron stars are discussed. It is argued that theoretical and experimental insights from the small baryo-chemical potential ($\mu_B$) and high-temperature regions of the QCD phase diagram can guide realistic model building at high density, as this regime is currently not accessible to first-principles numerical calculations of the QCD partition function. Special attention is payed to the chiral properties of high-density matter and the nature of a possible first-order chiral phase transition. In this transition hadronic parity-partners, in particular baryons, become spectrally degenerate with finite (pole) masses, as expected from general insight into the mass generation in QCD. Possible signals in heavy-ion dielectron production at beam energies of a few GeV are discussed. Based on evidence for an emergent "chiral spin symmetry" above the pseudo-critical chiral transition temperature at small $\mu_B$,  speculations on the  physical state of dense hadronic matter beyond the chiral phase transition are presented.           
\end{abstract}
\section{Introduction}
Hot and baryon-rich matter, governed by the strong interaction  (QDC matter), is encountered in astrophysical settings such the core-collapse supernova remnants shortly after bounce and the dynamics of binary neutron-star mergers. The latter have received much attention  through the observation of gravitational-wave radiation from two merger events, GW170817 and GW190425 by the LIGO-VIRGO collaboration~\cite{GW170817}\cite{GW190425}. In the post-merger phase densities of several times the saturation density of symmetric nuclear matter, $n_0=0.16$ fm$^3$, and temperatures of $T\sim 50-70$ MeV can be reached. In the laboratory such extreme conditions in temperature and density are created and diagnosed with relativistic heavy-ion collisions at center-of-mass energies of several GeV per nucleon pair. A sensitive probe is dielectron pair production which carries information about the early stages of the collision, in particular the temperature of the hot fireball. Recent measurements for {\it Au-Au} collisions at $\sqrt{s_{NN}}=2.42$ GeV by the HADES collaboration report values of $n/n_0\sim 3$ and
$T\sim 72$ MeV~\cite{HADES} .
 
In the following I discuss properties of high-density nuclear matter with special focus on aspects of chiral symmetry of the QCD Lagrangian in the light-quark sector; its spontaneous breaking in the vacuum and its restoration in QCD matter. Sect.~1 reviews pertinent properties of strong-interaction matter at small $\mu_B$ which I consider relevant for the high-density region of large $\mu_B$. It also contains a short digression of the origin of mass in QCD which motivates chiral models at high baryon density. Sect.~2 deals with the chiral properties of dense nuclear matter especially the possibility of a first-order chiral phase transition in density regions relevant for the inner core of cold neutron stars and binary neutron-star mergers. I also discuss perspectives for the detection of such a chiral transition in heavy-ion experiments at collision energies of a few GeV currently performed by the HADES collaboration and in the future at the FAIR facility in Darmstadt. Sect.~3 concludes with an outlook.   

\section{Lessons from high temperature and small baryo-chemical potential}
Most of our theoretical knowledge of the phase diagram of QCD matter at small $\mu_{B}$  is obtained from first-principles numerical evaluation for the QCD partition function on a discretized Euclidean space-time lattice (LQCD) from which the equation of state (EoS) and imaginary-time correlation functions can be computed. Continuum-extrapolated results for physical quark masses indicate a rapid change of degrees of freedom in a cross-over transition transition. At the same time spontaneously broken chiral symmetry is restored as can be inferred from the temperature dependence of the chiral condensate, $\ave{\bar qq}$, with a pseudo-critical temperature of $T^{pc}_{\chi}\sim 156$ MeV~\cite{Karsch} (and references therein). Although the Polyakov-loop expectation value - an order parameter for the deconfinement transition with infinitely heavy quarks - slowly reaches its deconfined value with increasing temperature~\cite{Petreczky} the nature of the deconfinement transition remains under debate. I shall return to this point below.

Experimental information on the low-$\mu_B$ and high-$T$ region of the QCD phase diagram is obtained from the collision of medium- and high-mass nuclei at RHIC and LHC energies~\cite{PBM}. Several observables point to the presence of a deconfined quark-gluon plasma (QGP) for temperatures much larger than $T^{pc}_\chi$. In the present context it noteworthy that $T_{chem}$ for the chemical freeze of hadrons for $\mu_B/T\sim 2$ has been mapped out with high precision. Remarkably, $T_{chem}$ in this range coincides with LQCD results for $T^{pc}_{\chi}(\mu_B)$ within errors for reasons that are not yet fully understood. For the following measurements of low-invariant-mass dilepton spectra for invariant masses of up to $\sim$ 2 GeV are of particular interest. Since the mean-free path of real and virtual photons is much larger than the size of the fireball they provide undisturbed information from the initial state to the kinetic freeze out at which hadrons free-stream to the detector. As the electromagnetic field couples to all charges one can deduce the in-medium properties of hadrons and the change in degrees of freedom across the transition window for chiral symmetry restoration.     

\subsection{Dileptons from heavy-ion collisions at high energy}
\begin{figure}[h]
\begin{minipage}{16pc}
\includegraphics[width=19pc]{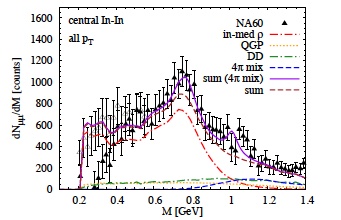}

\end{minipage}\hspace{4pc}
\begin{minipage}{17pc}
\includegraphics[width=17.5pc]{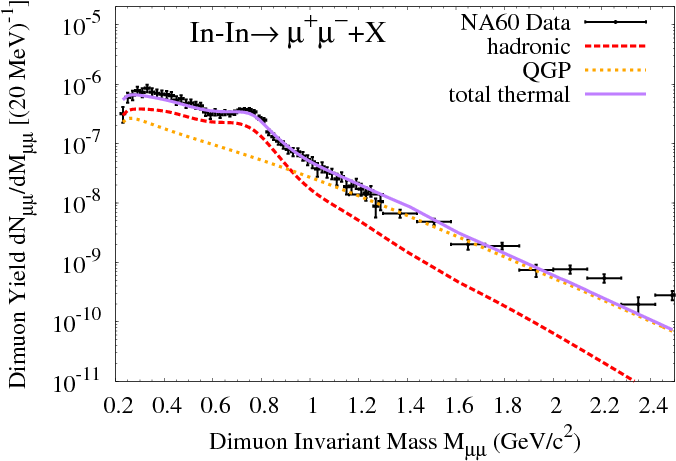}
\end{minipage}
\caption{The subtracted excess dimuon spectrum in the invariant-mass region of the $\rho$ meson (left panel) and the total excess yield (right panel) as measured by the NA60 Collaboration~\cite{Na60coll} is compared to theoretical results from Refs.~\cite{Trenton} (left, Fig.~from~\cite{Trenton}) and~\cite{Rapp Hees} (right, Fig.~from~\cite{Rapp Hees}).} 
\label{Fig:Na60}
\end{figure}

From measured $e^+e^-\!\!\to\rm{hadrons}$ annihilation cross sections for invariant masses of up to $\sim$ 1 GeV one knows that the rate is dominated by the $\rho$-meson resonance (with a small contribution from the $\omega$ meson). This is well understood from simple quark-charge counting in the electromagnetic current and leads to the vector-dominance model (VDM)~\cite{Sakurai}. A comparison of high-statistics 
dimuon measurements in central {\it In-In} collisions at the CERN-SPS by the NA60 collaboration~\cite{Na60coll} with a hadronic many-body calculation~\cite{RW}, supplemented by Hard-Thermal-Loop improved quark-antiquark annihilation~\cite{Braaten}, shows excellent agreement for the background subtracted rates in the $\rho$-meson region (left panel of Fig.~\ref{Fig:Na60})~\cite{Trenton} as well as the total subtracted yields (right panel of Fig.~\ref{Fig:Na60})~\cite{Rapp Hees}.

\noindent
Several conclusions can be drawn: 

1. A $\rho$ meson propagating in the thermal fireball suffers collisional broadening which becomes very large near and above $T^{pc}_{\chi}$ and merges smoothly into the $\bar qq$ continuum. The centroid of the invariant-mass distribution remains essentially temperature independent indicating that the $\rho$-meson pole mass is barely shifted. 

\begin{figure}[h]
\begin{minipage}{16pc}
\includegraphics[width=19pc]{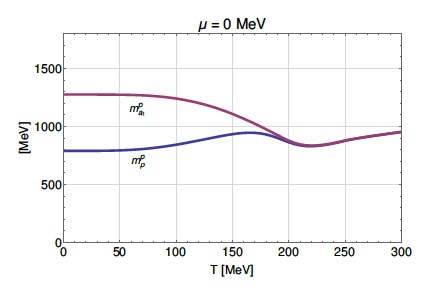}
\end{minipage}\hspace{4pc}
\begin{minipage}{17pc}
\hspace{1cm}\includegraphics[width=14pc]{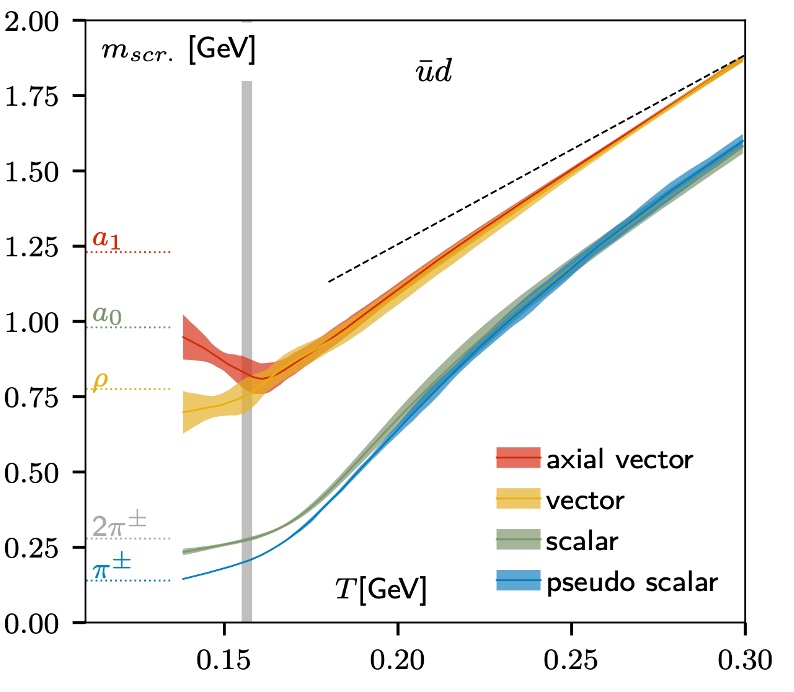}
\end{minipage}
\caption{Temperature dependence of  the $\rho$-  and $a_1$ pole masses~\cite{Jung} (left panel, Fig.~from~\cite{Jung}) and $\bar ud$ meson screening masses~\cite{Baza2} (right panel, Fig.~from~\cite{Baza2}).}
\label{Fig:rho-a1} 
\end{figure}

2. Above invariant dilepton masses of 1 GeV, effects from "chiral mixing", caused by admixtures of the in-medium $a_1$ meson (the parity partner of the $\rho$ meson) become important.  When chiral symmetry is restored they lead to nearly complete degeneracy of the $\rho$- and $a_1$ mass distributions as has been shown in~\cite{Hohler}. This finding was independently confirmed in a study of the in-medium $\rho$- and $a_1$ spectral functions (SpF's)~\cite{Jung} in a two-flavor quark-meson model which showed a merging of the pole masses after symmetry restoration (left panel of Fig.~\ref{Fig:rho-a1}). In fact, in this study it was established that the SpF's degenerate for all energies $p_0$ and three-momenta $\vec p$, as it should be. The results are corroborated by LQCD studies of temperature-dependent Euclidean (screening) masses of parity partners involving light $\bar ud$ quarks~\cite{Baza2} (right panel of Fig.~\ref{Fig:rho-a1}), thus firmly establishing "parity doubling" in the meson sector. Similarly, parity doubling in lattice studies of the screening masses of baryon octet and decuplet states~\cite{Aarts} was found. The latter is highly relevant for the high-$\mu_B$ region, as will be discussed below. 
\subsection{Emergent "chiral spin symmetry"}
The question arises what the physical nature of mesonic states after chiral restoration is. Do they rapidly dissolve into light quark-antiquark pairs, as expected from a weakly-coupled QGP, or do they show strong residual correlations? In this respect the suggestion of an emergent symmetry~\cite{Glozman1} is relevant (see~\cite{Glreview} for a recent review). In LQCD simulations of $J=1$ meson screening masses in which spontaneous chiral symmetry breaking was artificially suppressed by removing the lowest (near-zero) modes of the quark propagator, a new degeneracy pattern has been observed~\cite{Glozman2}. For $N_f$ quark flavors this pattern can be classified according to an $SU(2)_{CS}\times SU(2N_f)$ group called "chiral spin symmetry". It has been concluded that the confining chromo-electric interaction is distributed over all modes of the Dirac operator while the chromo-magnetic interaction mostly affects the near-zero modes. The chiral spin symmetry group is larger than the original chiral symmetry group $SU(N_f)_L\times SU(N_f)_R\times U(1)_A$ of the QCD Lagrangian which it contains as a subgroup. The larger symmetry is however a symmetry of the confining chromo-electric interaction (color charge) in a given frame. Thus it can only be (approximately) realized in situations where the electric quark-gluon interaction dominates over the magnetic one and kinetic terms. This seems to happen near $\mu_B=0$ in a temperature range $T^{pc}_{\chi}\lesssim T\lesssim 3T^{pc}_{\chi}$ as LQCD simulations suggest~\cite{Glozman3}\cite{Glozman4}. In this temperature range a physical picture emerges where configurations of nearly massless quark-anitquark pairs are bound by color-electric flux tubes into confined color singlets. As such configurations are reminiscent of a "string", this thermal state of QCD matter has been coined a "stringy fluid"~\cite{Glozman3}\cite{Glozman4}. For temperatures beyond $\sim 3T^{pc}_{\chi}$ the chromo-electric interaction between quark-antiquark pairs gets Debye-screened and the QGP state of deconfined, weakly-interacting quarks and gluons is gradually reached. The general conclusion is therefore that the chiral- and deconfinement transition do not coincide.
\subsection{Mass generation in QCD}
The fact that hadronic states retain most of their mass beyond the chiral cross-over temperature $T^{pc}_{\chi}$ warrants a brief discussion of the origin of mass in QCD. 
It is clear that the Higgs masses of light $u,d$ quarks only make a small contribution the nucleon mass $M_{N}$. The major part must arise from non-perturbative quantum effects through dimensional transmutation which renders a non-vanishing trace of the energy-momentum tensor (EMT)~\cite{EMT1}\cite{EMT2}:
\beq
\hspace{2cm}T^\mu_\mu=\frac{\beta(g)}{2g}G^{\mu\nu a}G_{\mu\nu}^a + \sum_{l=u,d,s} m_l(1+\gamma_{m_l})\bar{q}_l q_l~,
\label{traceanomaly}
\eeq
where  $\beta(g)$ is the QCD $\beta$-function with coupling constant $g$, and $\gamma_{m_l}$ is the anomalous mass dimension of a light quark with flavor $l=u,d,s$.\footnote{The heavy-quark contribution to the second term in the trace anomaly can be absorbed into $\beta(g)$.}  For a nucleon of four-momentum $p$  and spin $s$ with state vector $\ket{N(p)}$, one considers the matrix element
\beq
\bra{N(p_1)}T^\mu_\mu\ket{N(p_2)}=\biggl(\frac{M_N^2}{p_{01}p_{02}}\biggr)^{1/2}\bar u(p_1,s_1)u(p_2,s_2) G(q^2)~,
\label{gravFF}
\eeq
with momentum transfer $q^2=(p_1-p_2)^2$. It has been shown in Ref.~\cite{Kharzeev} that the forward matrix element ($q^2=0$) in the rest frame is identical to the mass that enters the Einstein equations for gravity, i.e. $G(0)=M_N$. Hence $G(q^2)$ is called the (scalar) "gravitational" form factor.

In the chiral limit of massless light quarks the EMT only contains the gluon term and one concludes that in this limit the mass of the nucleon entirely originates from gluons. The second term (sigma term) for physical light quark masses can be extracted from pion and kaon scattering data or from LQCD. It is found that it contributes $\sim$ 80 MeV or about 8\% to the total rest mass of the proton~\cite{Liu}. It is to be expected that similar arguments apply to mesons containing light quarks (except the pion) although a quantitive understanding in this sector is far less developed.  

The $q^2$-dependence of the gravitational proton form factor $G(q^2)$ contains information about the size of the mass distribution which, similar to the charge radius $R_c$, is extracted from the slope of $G(q^2)$ at $t=q^2=0$ as the "gravitational radius":
\beq
R^2_M=\frac{6}{M}\frac{dG}{dt}\Big|_{t=0}~.
\eeq
$R_M$ can be measured in near-threshold photoproduction of quarkonium $J/\psi$ or $\Upsilon$ states. In the $t$-channel at large invariant mass predominantly two highly correlated gluons couple to the heavy $c\bar c$ or $b\bar b$ quarks. From data obtained in exclusive $J/\psi$ photoproduction on the proton by the GlueX Collaboration~\cite{GlueX} a value $R_M=0.55\pm0.03$ fm has been extracted in~\cite{Kharzeev} which is significantly small that the charge radius $R_c=0.8409\pm 0.0004$ fm~\cite{PDG}. Implications for dense nuclear matter will be discussed below.

\section{Dense nuclear matter}
The properties of dense hadronic matter at high $\mu_B$ and $T\lesssim 100$ MeV, which are reached in neutron-star mergers, are much less certain than those at low $\mu_B$. In part this is because the EoS cannot be studied from first principles LQCD because of the fermion sign problem. Currently one has to resort to effective theories involving baryon- and meson degrees of freedom. In constructing such theories one should be guided by the symmetry principles of QCD and knowledge from the small-$\mu_B$ region of the phase diagram, as discussed in the previous section. Both, spontaneously broken chiral symmetry and the trace anomaly should be incorporated in building appropriate effective Lagrangians.

 \subsection{The parity doublet model (PDM)}
As argued before there is evidence that chiral symmetry requires the treatment of hadronic parity partners on equal footing. The observation that hadrons remain massive across the chiral restoration transition should also be respected. Therefore, when dealing with the nucleon, its negative parity partner, the 
$N^*(1535)$, also has to be included as a fermionic degree of freedom\footnote{Numerical LQCD studies of the temperature dependence of the screening masses of the $N(939)$ and the $N^*(1535)$ at $\mu_B=0$ show that they become (nearly) equal at the pseudo-critical temperature $T^{pc}_{\chi}$ where the in-medium mass of the nucleon is essentially temperature independent~\cite{Aarts}.}.    

The problem of an effective chiral Lagrangian for massive fermions has been addressed in~\cite{DeTar}. Denoting $N_1$ and $N_2$ as the fields of the nucleon and its parity partner and introducing left- and right-handed fields as  $N_{iR,L}=(1\pm\gamma_5)N_i/2$, a "mirror assignment":
\bea
N_{1R}\to R N_{1R},\quad N_{1L}\to L N_{1L}
\label{Eq:mirror1}\\
N_{2R}\to L N_{2R},\quad N_{2L}\to R N_{1L}
\label{Eq:mirror2}
\eea	
is defined~\cite{DeTar}. The isospin transformations $L,R\in SU(2)_{L,R}$ act independently on the left- and right-handed components of the nucleon field, $N_1$, in Eq.~(\ref{Eq:mirror1}). On the other hand the right-handed component of the negative parity partner, $N_2$, transforms like a left-handed nucleon and vice versa (Eq.~(\ref{Eq:mirror2})). This construction ensures that the massive fermionic Lagrangian is chirally invariant. When supplemented by a mesonic Lagrangian involving $\sigma$- and pion fields as an $O(4)$ vector $\phi=(\sigma,\vec\pi)$ the simplest version of the PDM Lagrangian reads:  
\bea
	     {\cal L}_{\rm{PDM}}&=&\bar N_1\left (i\slashed\partial+g_1(\sigma+i\vec{\tau}\cdot\vec{\pi}\gamma_5\right)N_1 
	     +\bar N_2\left (i\slashed\partial+g_2(\sigma-i\vec{\tau}\cdot\vec{\pi}\gamma_5\right)N_2\nonumber\\
	     &-&m_{0,N}\left(\bar N_1\gamma_5N_2-\bar N_2\gamma_5N_1\right)+{\cal L}_{meson}\nonumber
\eea
\vspace{-0.6cm}
\beq	     
	    {\cal L}_{meson}=\frac{1}{2}\partial_\mu\sigma\partial^\mu\sigma+\frac{1}{2} \partial_\mu\vec\pi\partial^\mu\vec\pi+U(\phi^2)-c\sigma;\quad U(\phi^2)=\frac{\bar\mu}{2}\bigl(\sigma^2+\vec\pi^2\bigr)-\frac{\lambda}{4}\bigl(\sigma^2+\vec\pi^2\bigr)^2+\cdots
\label{Eq:PDM1}
\eeq
If the mass $m_{0,N}$ is set to zero the part that only involves nucleons is the Gell-Mann-L\'evi sigma model. The nucleon mass then arises entirely from spontaneous chiral symmetry breaking and hence vanishes in the restored phase. One the other hand in the PDM it remains finite. According to the trace anomaly the mass parameter $m_{0,N}$ is of purely gluonic origin while the sigma terms are generated dynamically through spontaneous chiral symmetry breaking. 
 
\subsection{Phase diagram} 
One of the principal aims is to calculate the thermodynamic properties of nuclear matter such as the EoS, the velocity of sound, the specific heat etc. at large $\mu_B$ and finite $T$. Usually the mean-field approximation is used in which the $\sigma$ field in the PDM Lagrangian is replaced by its thermal expectation value. A more general approach is the Functional Renormalization Group method~\cite{Dupuis} which includes both thermal and quantum fluctuations. The starting point is the Euclidean effective action
\beq 
\Gamma(T,\mu_B)=\int_0^{1/T}\!\!\!\!d\tau\int \!d^3x \biggl({\cal L}_{PDM}^E-\mu_B\sum_{i=1,2}\bar N_i\gamma_0 N_i\biggr) .
\eeq
Upon regularization with a regulator function $R_k$ the effective action becomes scale dependent with momentum $k$ and its evolution is governed by a flow equation~\cite{Wetterich}
\beq
\partial _{k}\Gamma _{k}=\frac{1}{2}{\rm STr} \left(\partial_{k}R_{k} \left[ \Gamma_{k}^{(2)}+R_{k}\right]^{-1}\right),
\eeq 
where STr denotes the trace over space and all fermionic and bosonic indices and $\Gamma_{k}^{(2)}$ is the second functional derivative wrt to the boson - and fermion fields. The flow equation evolves 
$\Gamma_k$ from the classical action at some cut-off scale $k=\Lambda$ to the full quantum action, 
$\Gamma_{k=0}$, at the infrared scale $k=0$. $\Gamma_{k=0}$ is equivalent to the thermodynamic potential $\Omega(T,\mu_B)$ in the grand canonical ensemble. Thus the EoS and other thermal quantities can be computed and possible phase transitions identified. To lowest order in a gradient expansion of $\Gamma_k$ only the potential term $U_k(\phi^2)$ is scale dependent for which results will be presented.

\begin{figure}[h]
\begin{minipage}{16pc}
\includegraphics[width=17pc]{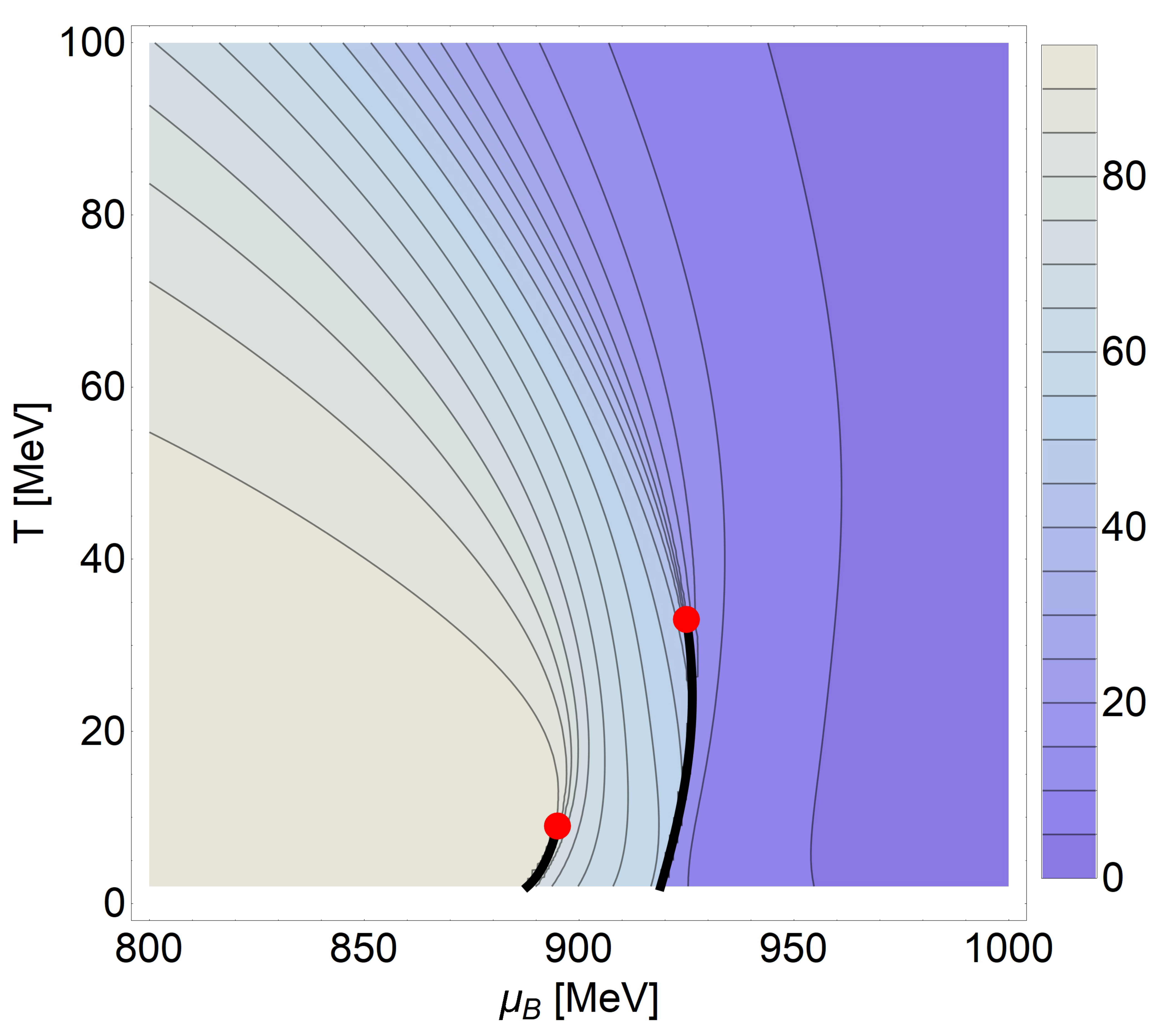}
\end{minipage}\hspace{4pc}
\begin{minipage}{17pc}
\hspace{0.5cm}\includegraphics[width=19pc]{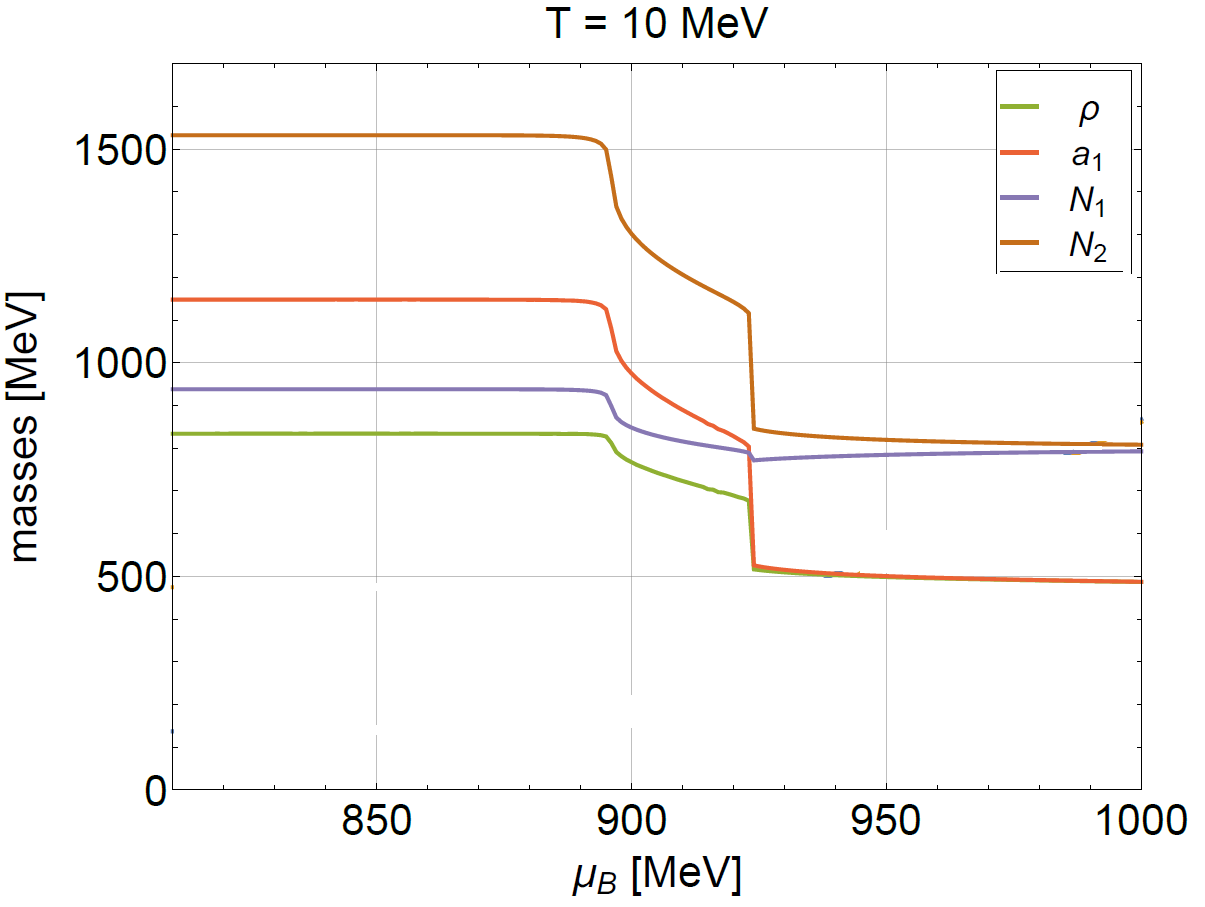}
\end{minipage}
\caption{Contour plot of the chiral condensate in the $T-\mu_B$ plain (left panel) and the 
$\mu_B$-dependent screening masses for the $N(939)$ and $N^*(1535)$ and the $\rho$- and $a_1$ meson (right panel) at $T=10$ MeV as calculated in Ref.~\cite{Trip1}. (Figs.~from~\cite{Trip1}).} 
\label{Fig:PDM1}
\end{figure}

The PDM phase diagram, i.e. the chiral condensate $\sigma(T,\mu_B)$, displayed in the left panel of Fig.~\ref{Fig:PDM1}~\cite{Trip1}, exhibits two distinct first-order phase transitions: a liquid-gas transition at smaller $\mu_B$ and a chiral transition at larger $\mu_B$. In the latter $\sigma(T,\mu_B)$ vanishes. The corresponding screening masses of the nucleonic parity partners $N_1$ and $N_2$ are modified at both transitions and become (almost) degenerate above the chiral transition. The same happens for the $\rho$- and $a_1$-meson masses. These $J=1$ mesons will be discussed further in the next section.  

One thus concludes that the PDM captures the essential features of mass generation in QCD in that hadron masses only partially result from spontaneously broken chiral symmetry and the strong $(T,\mu_B$)-dependence of the screening masses arises from chiral symmetry restoration through the in-medium sigma terms.  
At this point it should be mentioned that the chiral critical point (CEP) is highly sensitive to the bosonic quantum fluctuations of the $\sigma$- and pion fields. When these are artificially switched off in the FRG flow equations the chiral CEP ($\mu^c_{B\chi}=924$ MeV, $T_\chi^c=33$ MeV)  moves substantially upwards to $\mu^c_{B,\chi}\sim 1000$ MeV and $T^c_\chi\sim 120$ MeV~\cite{Trippriv}. 

\subsection{Is there an emergent chiral spin symmetry at large $\mu_B$?}
One can speculate whether the extended $SU(2)_{CS}\times SU(4)$ for $N_f=2$ ($u,d$) quarks which is likely to emerge at small $\mu_B$ beyond $T^{pc}_{\chi}$~\cite{Glreview} is also present at large $\mu_B$~\cite{GPP}.  In Ref.~\cite{Glozman5} it has been shown that the fermionic part of the PDM Lagrangian without coupling to the pion- and $\sigma$ field transforms under the chiral spin symmetry. Thus for this symmetry to be emergent beyond the first-order chiral transition (Fig.~\ref{Fig:PDM1}) the meson fields have to (approximately) decouple from the parity-doubled $N(939)$ and $N^*(1535)$ baryons. Should this happen, the physical picture of the new high-density phase might be similar to the small 
$\mu_B$ "stringy fluid" in that three light $u,d$ quarks are bound by color-electric flux tubes in confined baryons. These probably have a size of the order of the gravitational mass radius $R_M\sim 0.55$ fm, as opposed to the charge Radius of $R_c\sim 0.84$ fm and thus would occupy a much smaller volume. As a consequence, confined baryonic matter could persist to densities much higher than that of the chiral transition density. Admittedly, these scenarios are rather speculative at present and more work is needed to fully understand the physics of the high-$\mu_B$ phase. 

\subsection{Is parity doubling of the $N(939)$ and the $N^*(1535)$ observable?}
The question arises whether the degeneracy of the $N(939)$ and the $N^*(1535)$ in a chiral restoration transition at large $\mu_B$ could be observed. A promising experiment is the detection of dielectrons at low invariant masses and collision energies $\sqrt{s_{NN}}$ of a few GeV. To predict possible signals the PDM Lagrangian in Eq.~(\ref{Eq:PDM1}) has been extended to include $\rho$- and $a_1$-vector fields~\cite{Trip1}, as the virtual photon strongly couples to the $\rho$ meson and chiral mixing requires also the $a_1$ meson\footnote{ A novel  FRG formulation for massive vector fields based on (anti-)self-dual field strength~\cite{Jung2} has been used to avoid known problems of the Proca and St\"{u}kelberg constructions.}. Results for the $\mu_B$-dependent $\rho$ and $a_1$ screening masses are shown in the right panel of Fig.~\ref{Fig:PDM1}. It is interesting to see that both are again already modified at the liquid-gas transition and become nearly identical at the chiral first-order transition at a finite mass of $\sim$ 500 MeV. 

\begin{figure}[h]
\begin{minipage}{16pc}
\includegraphics[width=17.2pc]{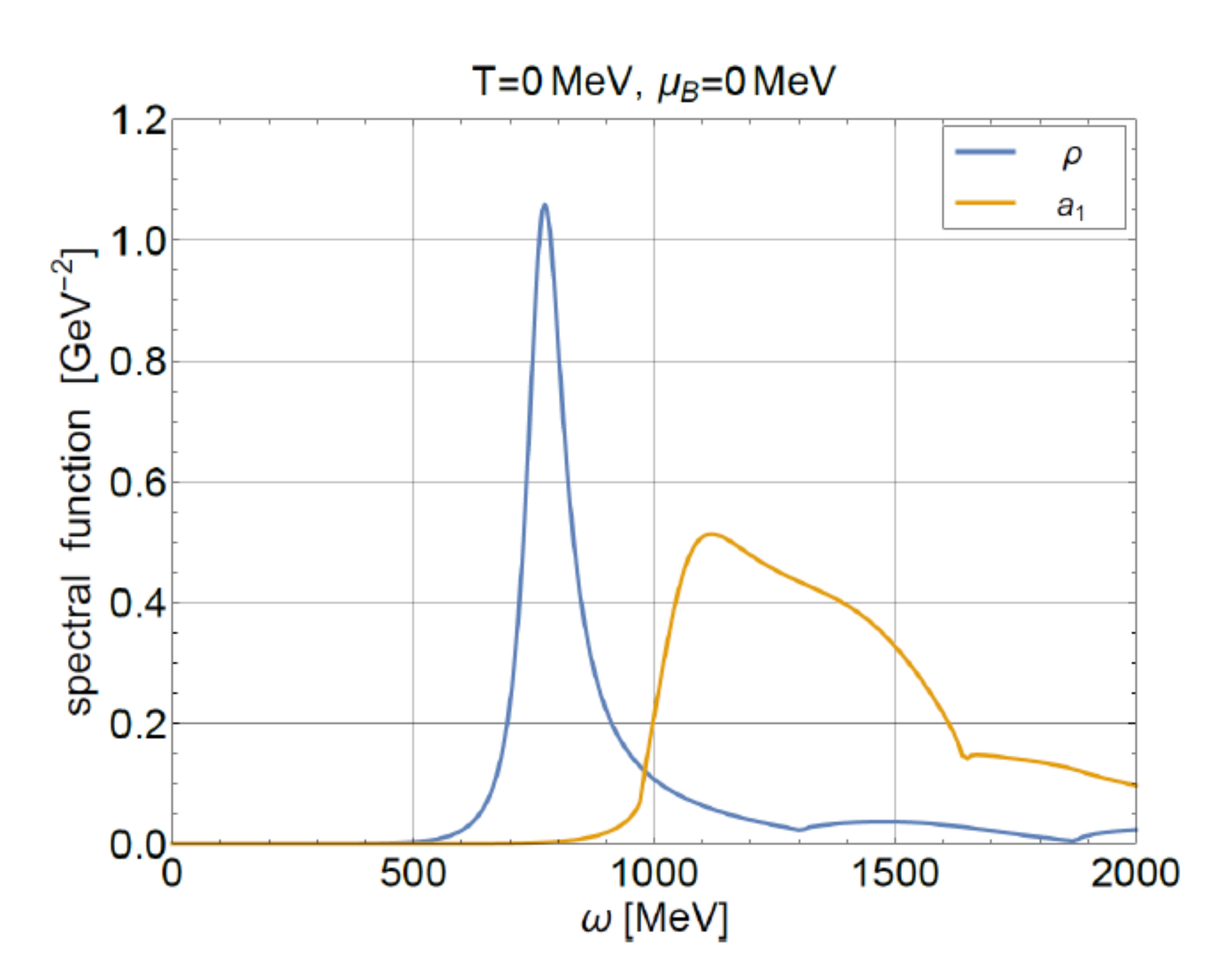}
\end{minipage}\hspace{4pc}
\begin{minipage}{17pc}
\hspace{0.0cm}\includegraphics[width=17pc]{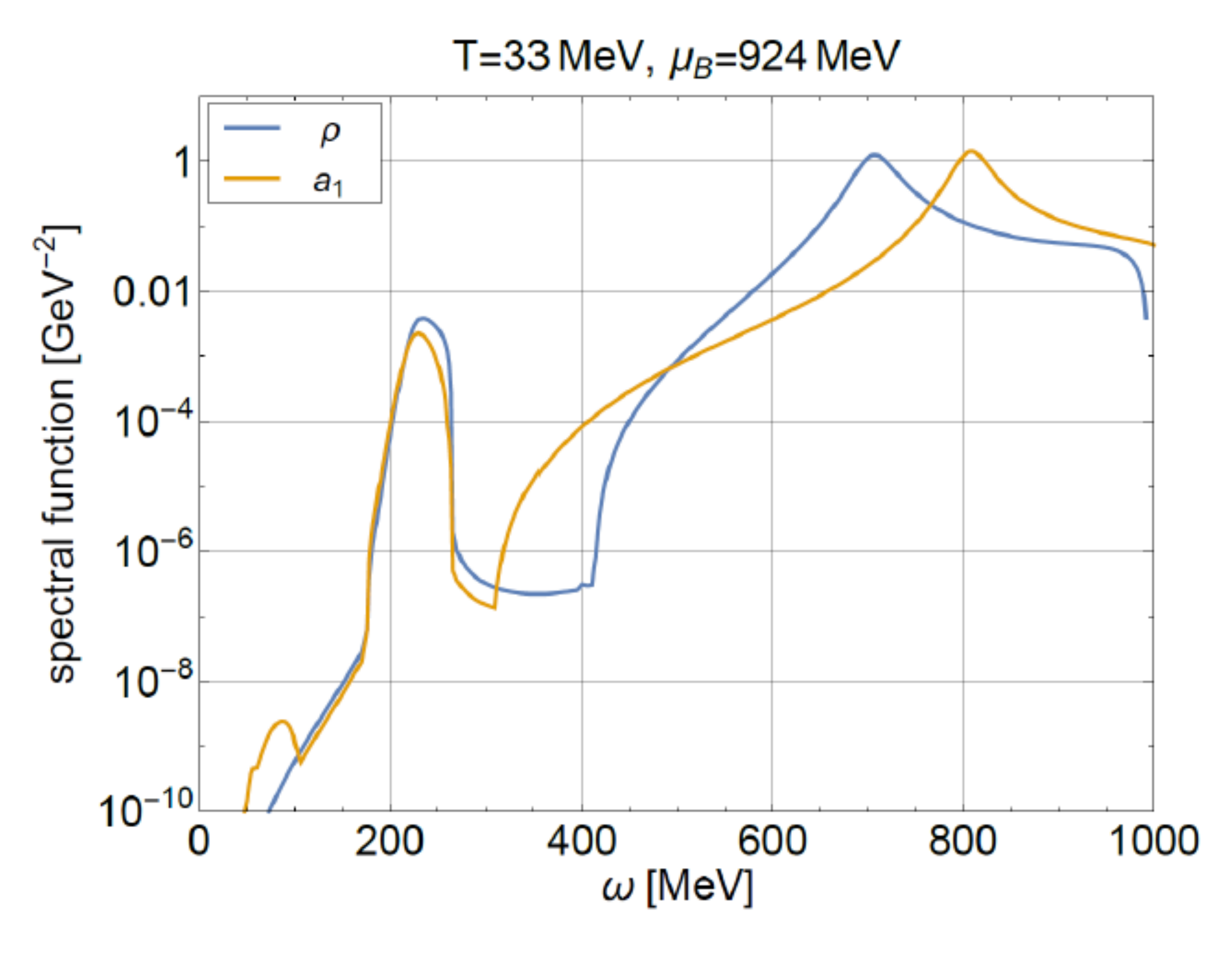}
\end{minipage}
\caption{Spectral functions for the $\rho$- and $a_1$ mesons in the vacuum (left panel) and near the chiral CEP (right panel) at vanishing 3-momentum $\vec p=0$ (Fig.~from~\cite{Trip1}).} 
\label{Fig:PDM2}
\end{figure}

For the calculation of dilepton production rates, the real-time $\rho$- and $a_1$ SpF's are needed. Within the FRG treatment of the extended PDM they can be obtained via analytic continuation of the corresponding Euclidean two-point correlation functions $\Gamma^{(2),E}_k$~\cite{Kamikado}~\cite{Trip2}. This leads to flow equations for the real and imaginary parts of the retarded correlation function $\Gamma^{(2),R}_k$  which are solved numerically for real frequencies $\omega\equiv p_0$ and three-momenta $\vec p$. In the infrared limit, $k\to 0$, the SpF's of the $J=1$ mesons $i$ are then given by
\beq
\rho_i(\omega,\vec p)=\frac{1}{\pi}\frac{\rm {Im}\Gamma^{(2),R}_i(\omega,\vec p)}{\bigl(\rm{Re}\Gamma^{(2),R}_i(\omega,\vec p)\bigr)^2+\bigl(\rm {Im}\Gamma^{(2),R}_i(\omega,\vec p)\bigr)^2};\quad i=\rho,a_1~. 
\eeq     
Results for $\vec p=0$ are shown in Fig.~\ref{Fig:PDM2}~\cite{Trip1}. 
In the vacuum the $\rho$-meson SpF shows a prominent peak at $\approx 775$ MeV while for the $a_1$ meson one observes a broader structure (left panel of Fig.~\ref{Fig:PDM2}). Close to the chiral CEP ($\mu^c_{B\chi}=924$ MeV, $T_\chi^c=33$ MeV) the SpF's of the parity partners become almost degenerate (right panel of Fig.~\ref{Fig:PDM2}) and exhibit a low-energy peak around $\omega\approx250$ MeV. Its origin can be traced back to the strong decrease of the in-medium $N(939)$ - $N^*(1535)$ mass gap, favoring low-energy baryon-resonance formation processes $\rho+N_1\to N_2$ and $a_1+N_1\to N_2$. This is a unique prediction of the baryonic mirror assignment.

\begin{figure}[h]
\begin{center}
\includegraphics[width=20pc]{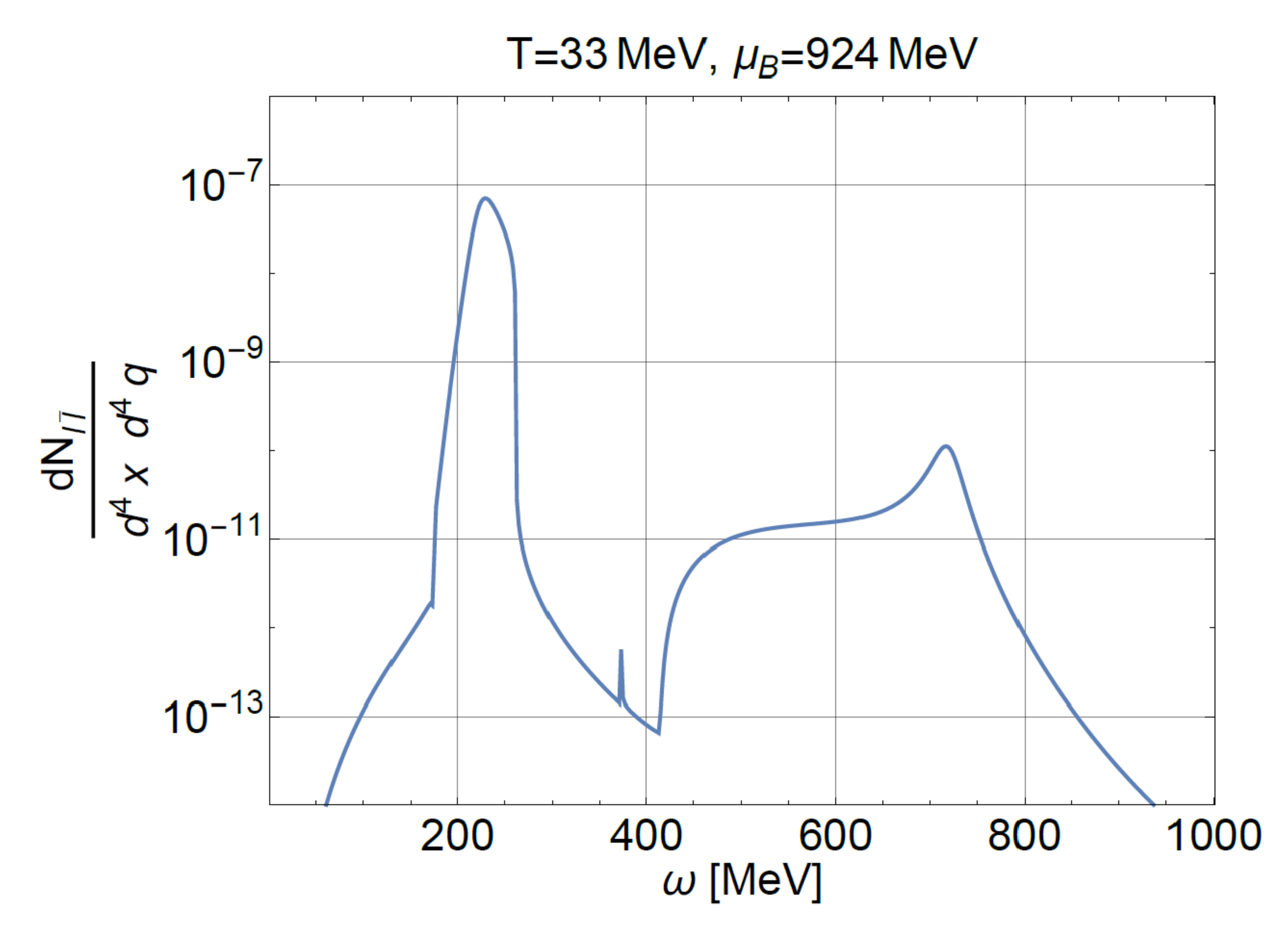}
\end{center}
\caption{Preliminary result for the invariant-mass-distribution of the thermal dielectron rate using the $\rho$-meson spectral function near the chiral CEP displayed in the right panel of Fig.~\ref{Fig:PDM2}. (Fig.~from~\cite{Trip4}).}
\label{Fig:Dilee}
\end{figure}

Since photons strongly couple to the $\rho$ meson the low-energy peak contributes to the dielectron yield at HADES energies as a low-invariant-mass enhancement of the rate. A preliminary estimate for the thermal dielectron rate, $dN_{e^+e^-}/d^4xd^4q$, near the chiral CEP, using the $\rho$-meson SpF in the right panel of Fig.~\ref{Fig:PDM2} has been obtained in~\cite{Trip4} and the result is displayed in Fig.~\ref{Fig:Dilee}.  
Given the fact that the measured $e^+e^-$- excess rate for invariant masses between $\sim150-800$ MeV is (nearly) exponential~\cite{HADES} and no peak is observed around 250 MeV the prediction in
Fig.~\ref{Fig:Dilee} does not seem to be realistic as additional collisional broadening from other hadron species in the fireball are not taken into account. Nonetheless a broader enhancement may be observable~\cite{Lari}. 
  
\section{Outlook}
From the arguments made above I believe that there is a case for a first-order chiral transition at high baryon density in which parity doubling of baryons and mesons occurs. Although this cannot be shown rigorously through first-principles LQCD calculations at present, chiral effective models which incorporate theoretical and experimental knowledge from the low-$\mu_B$ and high-$T$ region of the QCD phase diagram and the mass generation in QCD suggest a rich picture. Part of it is the $N(939)-N^*(1535)$ and 
$\rho-a_1$ parity degeneracy in the chiral transition which may be observable in heavy-ion experiments with virtual photons. The model studies also let one speculate about the properties of baryonic matter above the chiral transition. Admittedly the PDM results presented above are still qualitative. To be of relevance in astrophysical settings such as the inner core of neutron stars, core-collapse supernovae or neutron-star mergers they have to be quantified. One has to ensure that the thermodynamics obtained in chiral models with parity partners yields realistic EoS's for isospin-symmetric and -asymmetric nuclear matter that are compatible with observational constraints from cold neutron stars and agree with heavy-ion data. Work in this direction is in progress.        

\subsection{Acknowledgments}
I thank T. Galatyuk, C. Jung, L. von Smekal and R-A. Tripolt for a fruitful collaboration and many insightful discussions. This work was supported by the DFG Collaborative Research Centre 315477589-TRR 211, ”Strong interaction matter under extreme conditions”.


\section{References}
\begin{thebibliography}{100}
\bibitem{GW170817}
 Abbott B P  et al. (LIGO Scientific Collaboration and Virgo Collaboration) 2017 {\it Phys. Rev. Lett.} {\bf 119}
 161101 
\bibitem{GW190425}
 Abbott B P et al. (LIGO Scientific and Virgo Collaborations) 2020 {\it Astrophys. J. Lett.} {\bf 892} L3 
 \bibitem{HADES}
 The Hades Collaboration 2019 {\it Nat. Phys. {\bf 15}}  1040
 \bibitem{Karsch}
Bazavov A, et al. 2019  {\it Phys. Lett. B} {\bf 795} 15
\bibitem{Petreczky}
Petreczky P and H P Schadler 2015 {\it Phys. Rev. D} {\bf 92} 094517 
\bibitem{PBM}
Braun-Munzinger P, Rustamov A and Stachel J  2022  [arXiv:2211.08819 [hep-ph]] 
\bibitem{Sakurai}
 Sakurai J J 1960 {\it Annals of Physics} {\bf 11} 1
\bibitem{Na60coll}
Arnaldi A et al. (NA60 Collaboration) 2006 {\it  Phys. Rev. Lett.} {\bf 96}  162302
\bibitem{RW}
Rapp R and Wambach J 2000 {\it Adv. Nucl. Phys.} {\bf 25} 1
\bibitem{Braaten}
Braaten E, Pisarski R D and Yuan T C 1990 {\it Phys. Rev. Lett.} {\bf 64} 2242
\bibitem{Trenton}
Rapp R, van Hees H  and Strong T 2007 {\it Braz. J. Phys.} {\bf 37} 779
\bibitem{Rapp Hees}
Rapp R, van Hees H 2016 {\it Phys. Lett. B} {\bf 753} 586
\bibitem{Hohler}
Hohler P M and Rapp R 2014 {\it Phys. Lett. B} {\bf 731} 103
\bibitem{Jung}
Jung C, Rennecke F, Tripolt R-A, von Smekal L and Wambach J 2017 {Phys. Rev. D} {\bf 95} 036020
\bibitem{Baza2}
Bazavov A., et al. 2019 {\it Phys. Rev. D} {\bf 100} 094510
\bibitem{Aarts}
Aarts G, et al. 2019 {\it Phys. Rev. D} {\bf 99} 074503
 \bibitem{Glozman1}
Glozman L Y 2015 {\it Eur. Phys. J. A} {\bf 51} 27
\bibitem{Glreview}
 Glozman L Y 2022 [arXiv:2209.10235 [hep-lat]]
 \bibitem{Glozman2}
Denissenya M, Glozman L Y and Lang C, 2015 {\it Phys. Rev. D} {\bf 91} 034505
\bibitem{Glozman3}
Rohrhofer C et al, 2019 {\it Phys. Rev. D} {\bf 100} 014502
\bibitem{Glozman4}
Rohrhofer C, Aoki Y, Glozman L Y and Hashimoto S  2020 {\it Phys. Lett. B} {\bf 802} 135245
\bibitem{EMT1}
Collins J C, Duncan A S and Joglekar S D 1977 {\it Phys. Rev. D} {\bf 16} 438 
\bibitem{EMT2}
Nielsen N K 1977 {\it Nucl. Phys. B }{\bf 120} 212
\bibitem{Kharzeev}
Kharzeev D . 2021  {\it Phys. Rev. D} {\bf 104} 054015
\bibitem{Liu}
Liu K F 2021 {\it Phys. Rev. D} {\bf 104} 076010
\bibitem{GlueX}
Ali A et al. (GlueX Collaboration) 2019 {\it Phys. Rev. Lett.} {\bf 123} 072001 
\bibitem{PDG}
Zyla P et al. (Particle Data Group) 2020 {\it Progress of Theoretical and Experimental Physics} {\bf 2020} 083C01
\bibitem{DeTar}
DeTar C E and Kunihiro T 1989 {\it Phys. Rev. D} {\bf 39} 2805
\bibitem{Dupuis}
Dupuis N et al. 2021 {\it Physics Reports} {\bf 910} 1  
\bibitem{Wetterich}
Wetterich C 1993 {\it Phys. Lett. B} {\bf 301} 90
\bibitem{Trip1}
Tripolt R-A, Jung C, von Smekal L and Wambach J 2021 {\it Phys. Rev. D} {\bf 104} 054005
\bibitem{Jung2}
Jung C and  von Smekal L  2019 {\it Phys. Rev. D} {\bf 100} 116009
\bibitem{Trippriv} 
Tripolt R-A 2022 {\it private communication}
\bibitem{GPP}
Glozman L Y, Philipsen O and Pisarki R D 2022 [arXiv:2204.05083 [hep-ph]]
\bibitem{Glozman5}
Catillo M and Glozman L Y 2018 {\it Phys. Rev. D} {\bf 98} 014030
\bibitem{Kamikado}
Kamikado K, Strodthoff N, von Smekal L and Wambach J 2014 {\it Eur. J. Phys. C} {\bf 74} 2806
\bibitem{Trip2}
Tripolt R-A, Strodthoff N, von Smekal L and Wambach J 2014 {\it Phys. Rev. D} {\bf 89} 034010
\bibitem{Trip4}
Tripolt R-A and Geurts F,  2022 [arXiv:2210.01622 [hep-ph]]
\bibitem{Lari}
Larionov A B and von Smekal L, 2022 {\it Phys. Rev. C} {\bf 105} 034914 
\end{thebibliography}
\end{document}